\documentclass[aps,prl,preprint,superscriptaddress]{revtex4}

\usepackage{graphicx}
\usepackage{amsmath}
\usepackage{bm}

\begin{document}


\title{Neutralino dark matter scattering and 
$B_s \rightarrow \mu^+ \mu^-$ in SUSY models}



\author{S. Baek}
\email[]{sbaek@kias.re.kr}
\affiliation{Department of Physics, KIAS, Korea}

\author{Y.G. Kim}
\email[]{yg-kim@korea.ac.kr}
\affiliation{Department of Physics, Korea University, Seoul 136-701, Korea}

\author{P. Ko}
\email[]{pko@muon.kaist.ac.kr}
\affiliation{Department of Physics, KAIST, Daejon 305-701, Korea}


\date{\today}

\begin{abstract}
It is pointed out that there is a strong correlation between the neutralino 
dark matter scattering cross section $\sigma_{\tilde{\chi} p}$  and 
the branching ratio for $B_s \rightarrow \mu^+ \mu^-$ 
within minimal supergravity (mSUGRA) and its extensions. 
This correlation arises mainly from $\tan\beta$ and heavy neutral 
Higgs mass dependence, and shows a nice interplay between vastly different 
two observables within supersymmetric models. Current upper limit on 
$B( B_s \rightarrow \mu^+ \mu^- ) < 5.8 \times 10^{-7}$ excludes substantial 
parameter space where $\sigma_{\tilde{\chi} p}$ is within the CDMS  
sensitivity region. 
\end{abstract}


\maketitle


Minimal supersymmetric standard model (MSSM) is a well motivated candidate 
for physics beyond the standard model (SM). MSSM is consistent with precision 
electroweak data, and nicely complies with gauge coupling unification.  
Another nice feature of MSSM with $R-$parity conservation is the presence of 
natural candidates for cold dark matter (DM) of the universe. 
Recent data from WMAP collaboration indicates that 
$\Omega_{\rm DM} h^2 \simeq (0.095-0.13)$ \cite{Spergel:2003cb}, 
which could be dominated by the relic density of neutralino within SUSY 
models. In supergravity models, the LSP is the neutralino with 
$m_{\tilde{\chi}} \sim O( M_Z ) - O( G_F^{-1/2} )$ and could have suitable 
relic density. 

There has been experimental progress in direct detection of neutralino 
DM through (in)elastic scattering on various nuclei. Such experiments 
can be sensitive to a neutralino DM with mass $O(100)$ GeV, which is 
usually the case in various supergravity scenarios. 

Recently, the DAMA signal region \cite{Belli:2002yt} has been excluded
by the CDMS cryogenic  DM search experiment \cite{cdms} in the range of
\[
\sigma_{\tilde{\chi} p} = (10^{-6} - 10^{-5} )~{\rm pb},  
\]
with the corresponding DM mass depends on galactic halo models.
Since the CDMS experiment probes the DM scattering down to
$3 \times 10^{-7}$ pb level in certain range of DM mass,
it is important to calculate the DM scattering cross section within well
defined and/or motivated SUSY models, in which the cross section can be
in the CDMS sensitivity.

In Ref.s~\cite{Baek:2002rt}\cite{Baek:2002wm}, two of us considered a number 
of low energy phenomena such as $(g-2)_\mu$, $B\rightarrow X_s \gamma$, 
$B\rightarrow X_s l^+ l^-$ and $B_s \rightarrow \mu^+ \mu^-$ within various
SUSY breaking mediation mechanisms. In the present work, we extend our study 
to the neutralino DM scattering cross section $\sigma_{\tilde{\chi} p}$, its 
relic density $\Omega_{DM} h^2$,  and $B ( B_s \rightarrow \mu^+ \mu^- )$ 
in a class of (string inspired) supergravity models.  
We find that there is a strong correlation between
$\sigma_{\tilde{\chi} p}$ and  $B ( B_s \rightarrow \mu^+ \mu^- )$ for a 
given $\tan\beta$. The origin of this correlation resides in 
$\tan\beta$ and neutral Higgs boson masses ($m_{H}, m_A$)  
within a given (string inspired) supergravity scenarios. 
In particular, a large $\sigma_{\tilde{\chi} p}$ implies a large 
$B( B_s \rightarrow \mu^+ \mu^- )$, which may exceed the current upper 
limit on this process. Before proceeding, 
let us mention that there is an important difference between our previous 
works and the present work. In Ref.s~\cite{Baek:2002rt,Baek:2002wm}, we did 
not assume the neutralino LSP since there are ways of avoiding problems with
charged particle LSP. On the other hand, we assume that the LSP is the 
lightest neutralino in the present work, and consider their scattering with 
nuclei. 
Therefore SUSY contributions to various observables [ including the process 
$B ( B_s \rightarrow \mu^+ \mu^- )$ ] considered in this work are 
generically smaller than those given in Ref.s~\cite{Baek:2002rt,Baek:2002wm}. 

 

If the LSP is a neutralino of mass around $O( M_Z ) - O( v_{\rm EW})$, 
one can detect the relic neutralino LSP through (in)elastic scattering with 
various nuclei.  
In the large $\tan\beta$ limit, heavy neutral Higgs $H$ exchange contribution
to the DM scattering becomes important because of its  enhanced 
couplings to down type quarks such as strange or bottom quark. 
This is relevant to the heavy Higgs interaction with the strange quark 
contents inside nucleons, and the DM scattering cross section becomes 
enhanced. Therefore, the DM scattering amplitude increases linearly as 
$\tan\beta$ increases, and decreases as $m_A$ increases. 
Also the DM scattering amplitude is 
sensitive to the value of $\mu$, which determine the higgsino component 
of the neutralino because the neutralino-higgs coupling become significant 
when the neutralino is a mixed state of gaugino and higgsino.
We use the code DARKSUSY \cite{darksusy} in order to calculate the DM 
scattering cross section and its relic density within 
minimal supergravity with (non)universal Higgs mass parameters, 
and string inspired scenarios including a $D-$brane model.


The decay $B_s \rightarrow \mu^+ \mu^-$ can be an important probe of SUSY
in the large $\tan\beta$ limit, since its branching ratio grows like  
$\tan^6 \beta$ \cite{babu}. 
Unless the stop/charginos and neutral Higgs are too heavy, 
one can have a significant rate for this decay within SUSY models. If this 
decay is found at the level of $5 \times 10^{-7}$, then only gravity 
mediated SUSY breaking mediation (including string inspired scenarios) will 
survive (except for AMSB and no scale scenarios) 
\cite{Baek:2002rt,Baek:2002wm}.  Also, one can get a useful lower bound on 
$\tan\beta$, once this decay mode is observed \cite{kane}. 

In more general SUSY models where gluino mediated FCNC can be important, 
one has to include their effects and the correlation between the DM 
scattering cross section and the $B_s \rightarrow \mu^+ \mu^-$ branching 
ratio may be diluted. However gluino-mediated FCNC is not that important 
in the class of (string inspired) supergravity models we are considering, 
since the initial conditions for the soft parameters are universal or 
proportional to the Yukawa couplings, and $\delta$'s are generated mainly 
through RG evolution.

\begin{figure}
\includegraphics[height=5.0cm]{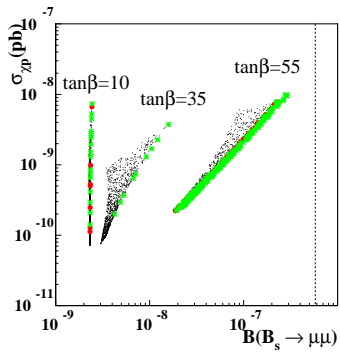}%
\caption{\label{fig1} 
$\sigma_{\tilde{\chi} p}$ vs. $B( B_s \rightarrow \mu^+ \mu^- )$ within 
mSUGRA with universal Higgs mass parameters for $\tan\beta = 10, 35 $ 
and 55 (from the left to the right). 
Black dots for $\Omega_\chi h^2 \geq 0.13$, 
red dots for $0.095 \leq \Omega_\chi h^2 \leq 0.13$ and
green dots for $\Omega_\chi h^2 \leq 0.095$.  
}
\end{figure}

The minimal SUGRA (mSUGRA) is specified by 5 parameters, 
\[
m_0~,~ m_{1/2}~,~ A_0~,~ \tan\beta~,~ {\rm sign} (\mu)~.
\]
The nature of the neutralino LSP is determined by gaugino mass parameters
$M_1$, $M_2$ and the $\mu$ parameter.  
$| \mu |$ is determined by the electroweak symmetry breaking condition: 
\begin{equation}
\mu^2 = {{m_{H_d}^2 - m_{H_u}^2 \tan^2 \beta } \over {\tan^2 \beta - 1 }}
- {1\over 2} M_Z^2 .
\end{equation}
In the mSUGRA scenario, $| \mu |$ is naturally large, so that the LSP is 
binolike and the (pseudo)scalar Higgs bosons $H$ and $A$ are  heavy.  
Therefore, the DM scattering cross section becomes small in this scenario, 
well below the CDMS  sensitivity region, and 
$B( B_s \rightarrow \mu^+ \mu^- )$  is not so much enhanced.
In Fig.~\ref{fig1},  we show the correlation between 
$\sigma_{\chi p}$ vs. $B( B_s \rightarrow \mu^+ \mu^- )$ within mSUGRA with 
$\tan\beta = 10, 35 $ and 55, respectively. For large $\tan\beta$, there is 
a strong correlation between the two observables, as emphasized in the 
beginning of this work. 
After imposing the $B\rightarrow X_s \gamma$ branching ratio as well as 
the lower bounds on the lightest Higgs mass and SUSY particle masses, and 
assuming the neutralino LSP, we find that the DM scattering cross section
is $\sigma_{\chi p} \lesssim 10^{-8}$ pb that is too small to be observed
at the current or near-future DM search experiemtns , and  
$B ( B_s \rightarrow \mu^+ \mu^- ) \lesssim 2 \times 10^{-7}$, 
which is below the reach of Tevatron. In particular, the current upper limit 
$B ( B_s \rightarrow \mu^+ \mu^- ) < 5.8 \times 10^{-7}$ \cite{cdf} 
does not put  any strong constraint on 
$\sigma_{\tilde{\chi} p}$ within the mSUGRA scenario
with universal Higgs mass parameters.

The universal soft parameters are too restricted assumption without solid
ground within supergravity framework. In order to consider more generic 
situation within supergravity scenario, let us relax the assumption of 
universal soft masses as follows: 
\begin{equation}
m_{H_u}^2 = m_0^2 ~( 1 + \delta_{H_u} ),~~~
m_{H_d}^2 = m_0^2 ~( 1 + \delta_{H_d} ),~~~
\end{equation}
whereas other scalar masses are still universal. 
Here $\delta$'s are parameters with $\lesssim O(1)$. This assumption is 
still too restrictive for the purpose of studying FCNC such as 
$B_s \rightarrow \mu^+ \mu^-$ within supergravity framework. 
On the other hand, the nonuniversality in the squark masses is not so 
important to the DM scattering, since in DM scattering what matters is 
the nature of the LSP, whether it is bino like or Higgsino like.  
The strong correlation between $\sigma_{\tilde{\chi} p}$ and 
$B ( B_s \rightarrow \mu^+ \mu^- )$ could be diluted if we allow
more general flavor structures in soft terms, which is visible in the
$D-$brane models we consider in this work. 

In order to emphasize the role of $B( B_s \rightarrow \mu^+ \mu^- )$, 
we take the numerical values of $\delta$'s as in Refs.~
\cite{Munoz:2003wx,Cerdeno:2003yt}: 
\begin{eqnarray}
(I) &  \delta_{H_d} = -1 , &  \delta_{H_u} = 0 , 
\nonumber  \\
(II) &  \delta_{H_d} = -1 , &  \delta_{H_u} = 1 .
\end{eqnarray}

\begin{figure*}
\begin{tabular}{cc}
{\includegraphics[height=5.0cm]{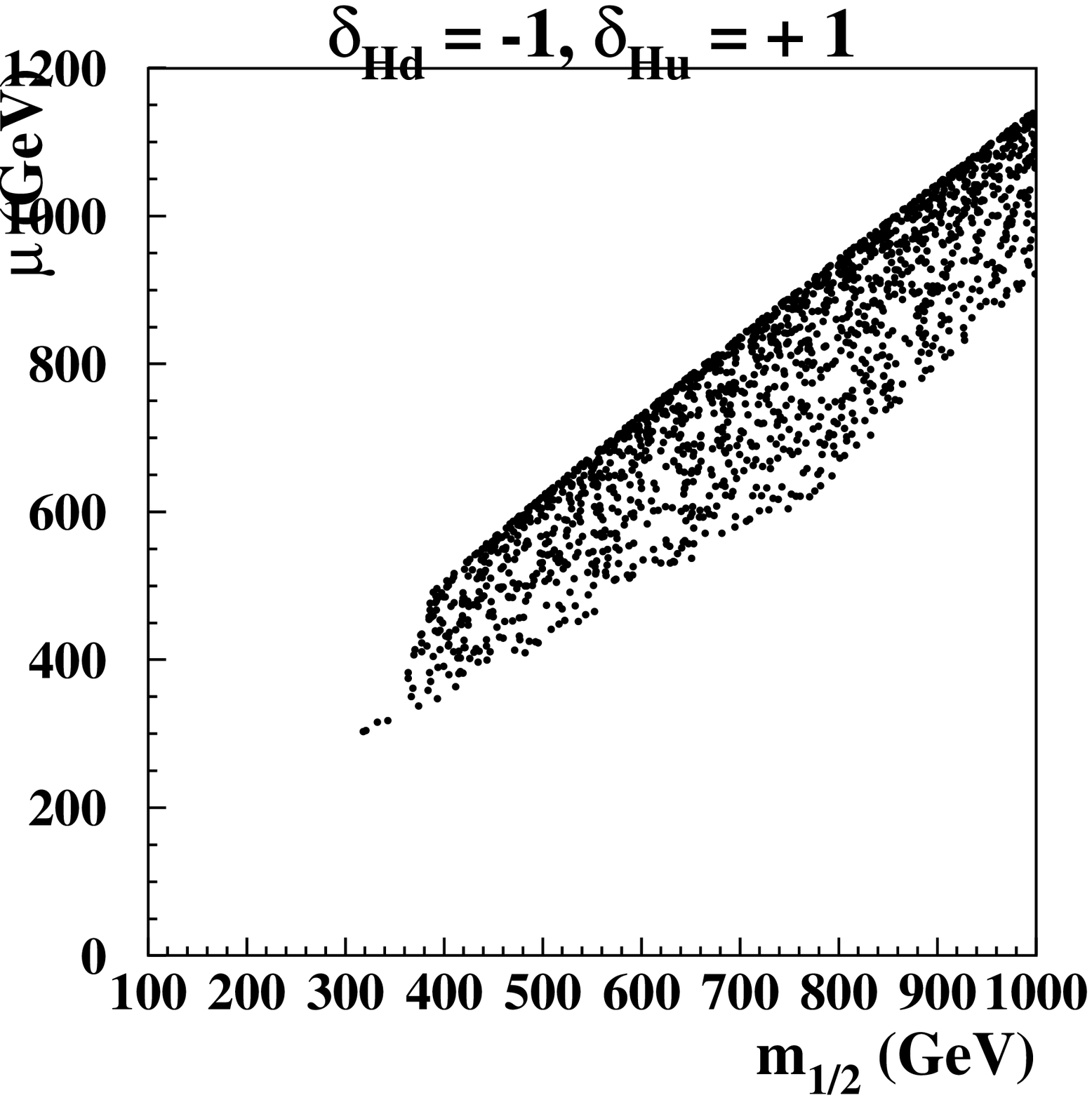}}&
{\includegraphics[height=5.0cm]{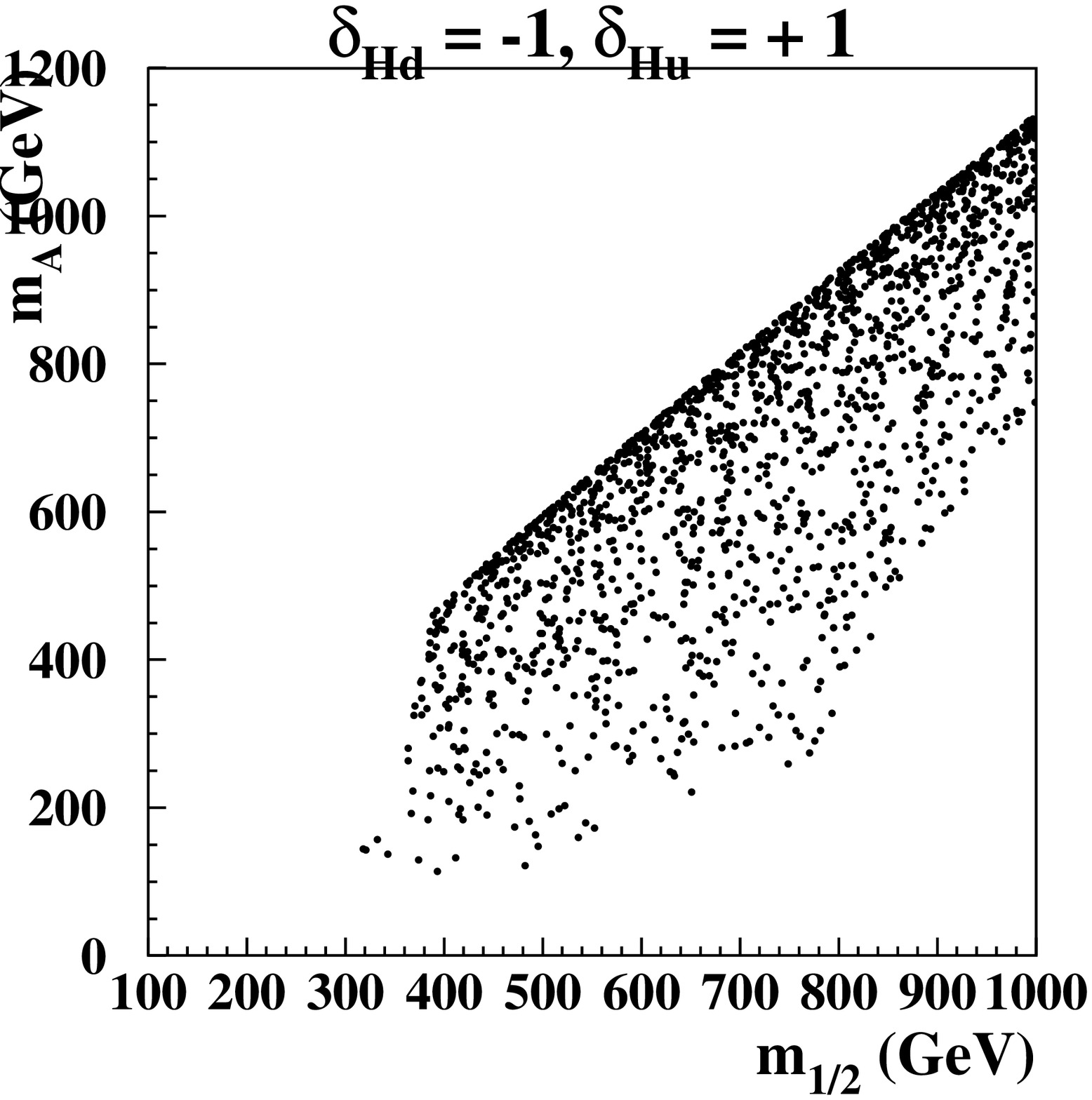}}
\end{tabular}
\caption{\label{fig2} 
(a) $\mu$ and (b) $m_A$ vs. $m_{1/2}$ in mSUGRA with nonuniversal Higgs 
mass parameters: $\delta_{H_u} = 1$ and $\delta_{H_d} = -1$.
}
\end{figure*}

In Fig.~\ref{fig2} (a) and (b), we show $\mu$ and the pseudoscalar mass
$m_A$ as functions of $m_{1/2}$ for the case (II). 
For $\delta_{H_u} = +1$, $\mu$ becomes lower and the Higgsino component
in the neutralino LSP increases so that $\sigma_{\chi p}$ is enhanced, as
discussed in Ref.~\cite{Munoz:2003wx}. 
The change of $|\mu|$ also has an impact on the higgs masses because
\[
m_A^2 = m_{H_u}^2 + m_{H_d}^2 + 2\mu^2 \simeq m_{H_d}^2 + \mu^2 - M_Z^2/2
\] 
at weak scale.  For $\delta_{H_d} = -1$, $m_A$ and 
$m_H$ becomes  further  lower, and 
both $\sigma_{\tilde{\chi}p}$ and $B( B_s \rightarrow \mu^+ \mu^- )$ are 
enhanced compared with the mSUGRA case. 
Note that the $B ( B_s \rightarrow \mu^+ \mu^- ) < 5.8 \times 10^{-7}$ 
provides a very significant constraint on the neutralino DM scattering cross
section $\sigma_{\tilde{\chi} p}$, and removes the parameter space where 
the DM scattering is within the reach of CDMS  experiment.

\begin{figure*}
\begin{tabular}{cc}
{\includegraphics[height=5.0cm]{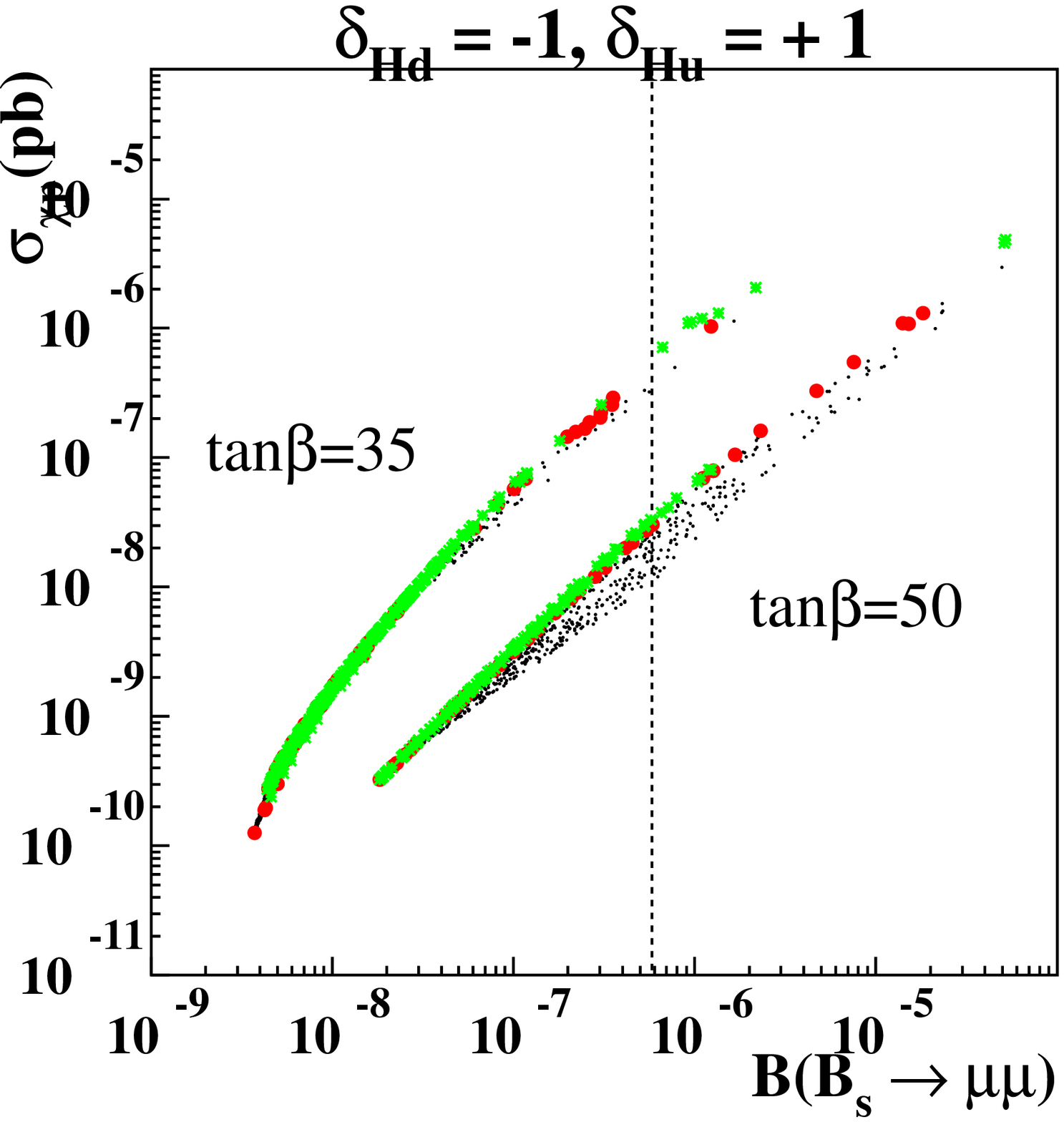}}&
{\includegraphics[height=5.0cm]{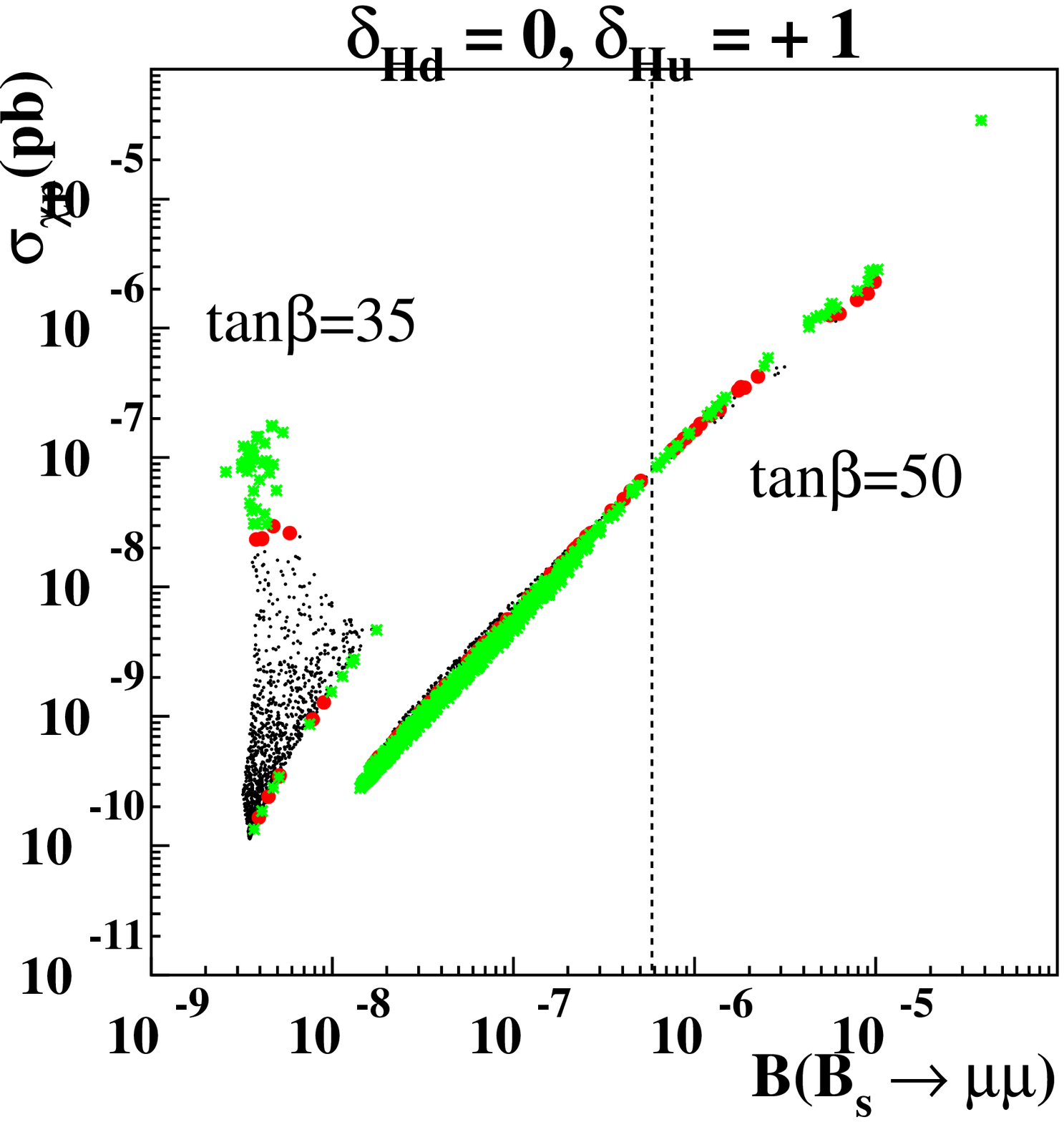}}
\end{tabular}
\caption{\label{fig3} 
$\sigma_{\tilde{\chi} p}$ vs. $B( B_s \rightarrow \mu^+ \mu^- )$ 
in mSUGRA with nonuniversal Higgs mass parameters: 
(a) $\delta_{H_u} = 1$ and $\delta_{H_d} = -1$ and 
(b) $\delta_{H_u} = 1$ and $\delta_{H_d} = 0$. 
}
\end{figure*}

\begin{figure*}
\begin{tabular}{cc}
{\includegraphics[height=5.0cm]{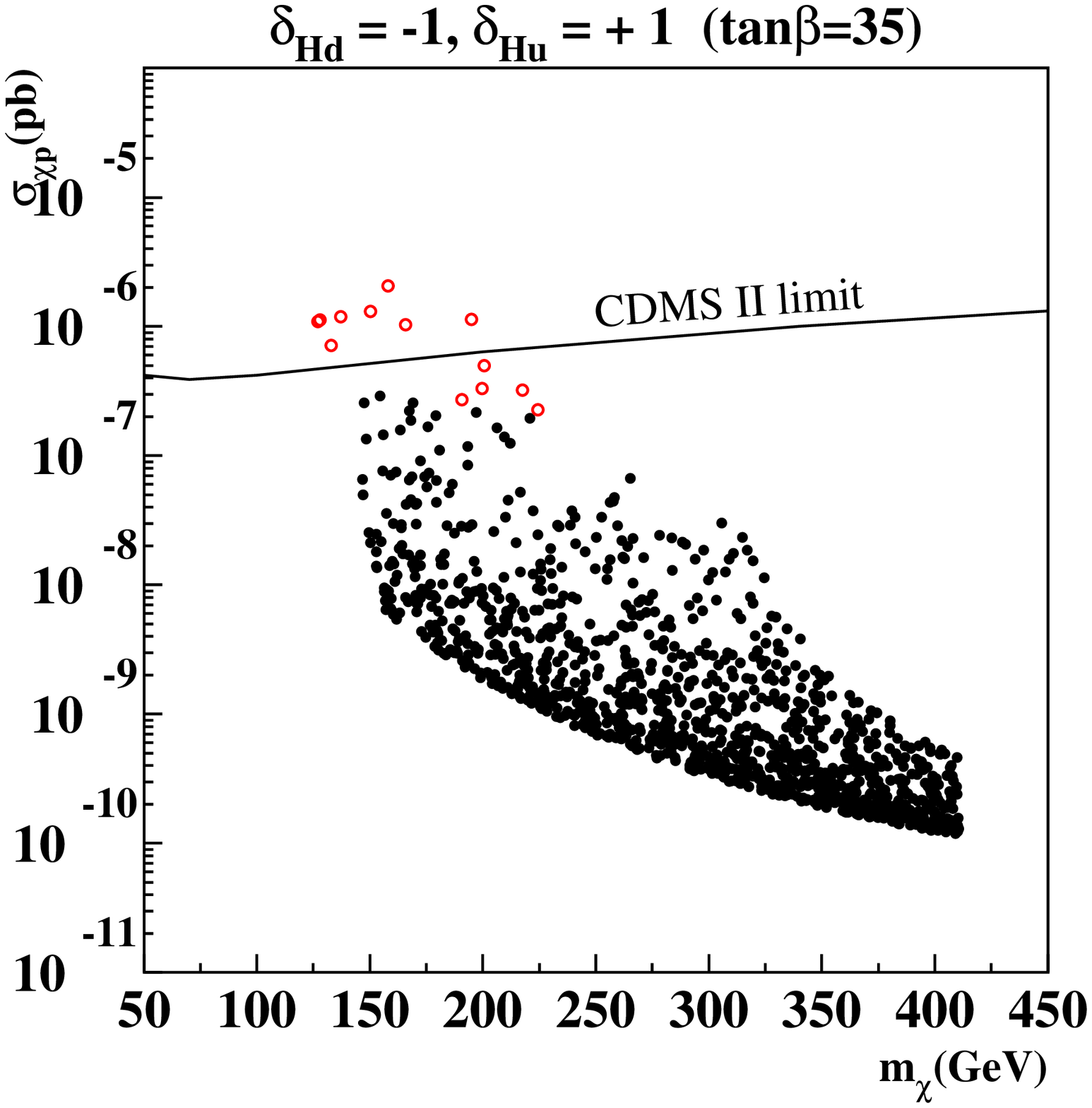}}&
{\includegraphics[height=5.0cm]{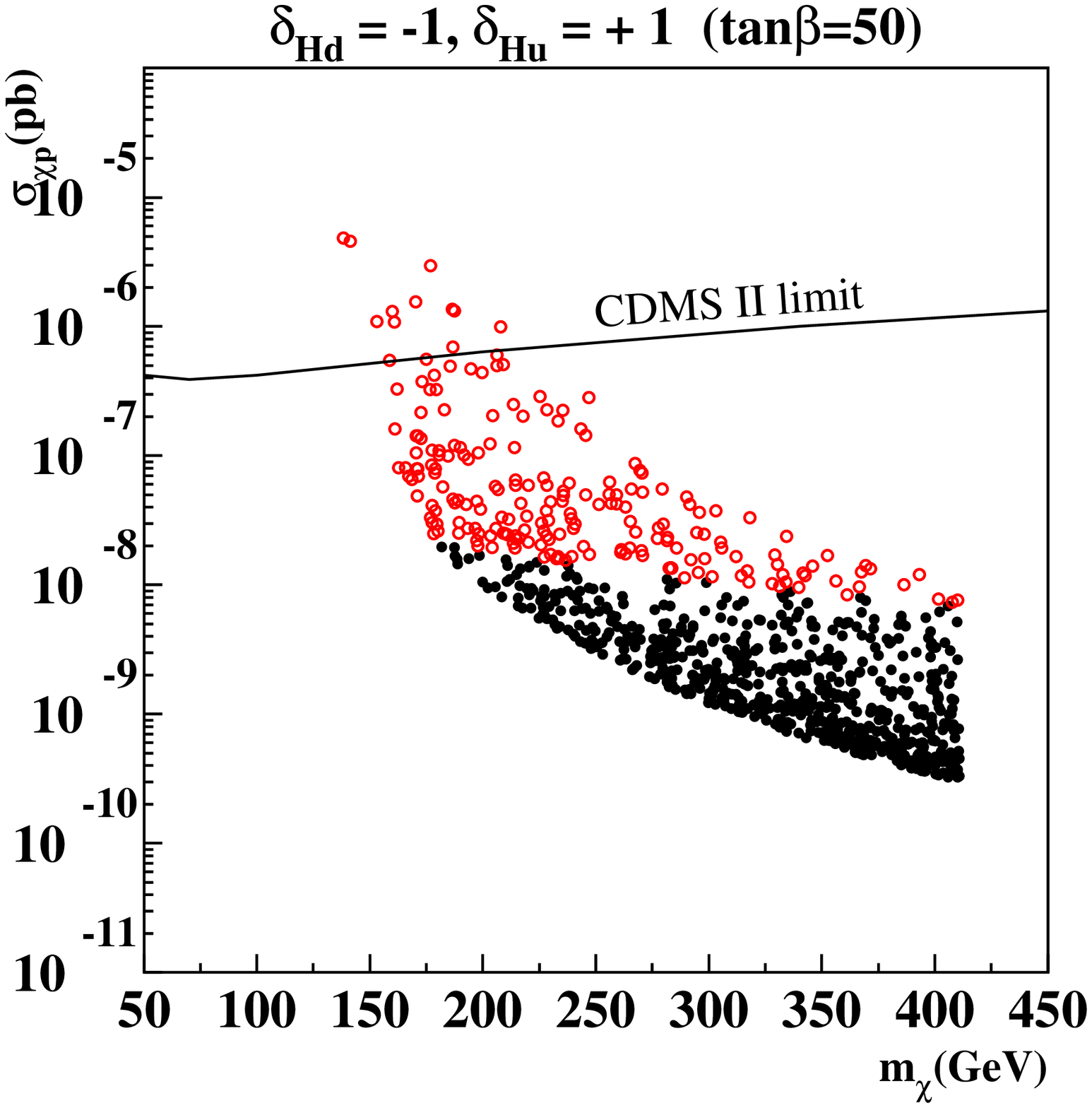}}
\end{tabular}
\caption{\label{fig32} 
$\sigma_{\tilde{\chi} p}$ vs. $m_{\chi}$  
in mSUGRA with nonuniversal Higgs mass parameters: 
$\delta_{H_u} = 1$ and $\delta_{H_d} = -1$ for
(a) $\tan\beta = 35$ and 
(b) $\tan\beta = 50$. 
The red points (open circles) are excluded by 
$B ( B_s \rightarrow \mu^+ \mu^- )$  constraint.}
\end{figure*}

In Fig.~\ref{fig32}, we show the scattered plot in the $( m_{\chi^0} , 
\sigma_{\chi p} )$ for $\delta_{H_d} = -1, \delta_{H_u} = +1$ along with 
the CDMS data for (a) $\tan\beta = 35$ and (b) $\tan\beta = 50$. 
Note that the constraints from the CDMS experiment and the 
$B ( B_s \rightarrow \mu^+ \mu^- )$  are comparable for $\tan\beta = 35$. 
However,  $B ( B_s \rightarrow \mu^+ \mu^- )$ becomes stronger   
for $\tan\beta = 50$. After imposing the $B ( B_s \rightarrow \mu^+ \mu^- )
< 5.7 \times 10^{-7}$ constraint for the $\tan\beta = 50$ case, 
we find that $\sigma_{\chi p} \lesssim 2 \times  10^{-8}$ pb, 
which is well below the current or near-future DM search experiments. 

We also considered nonuniversal gaugino masses, in which case the most 
important one is the gluino mass parameter via RG running. Therefore we 
considered a possibility that gluino mass can differ from the wino and bino
masses ($M_1 = M_2 \neq M_3$).   
We find that the qualitative  feature is similar to the case with  
nonuniversal Higgs masses.  In particular the current limit on 
$B ( B_s \rightarrow  \mu^+ \mu^- )$ already puts a strong constraint on 
$\sigma_{\tilde{\chi} p}$ in the large $\tan\beta$ region.


Next, we consider a specific $D$ brane model where the SM gauge groups 
and 3 generations live on different $Dp$ branes \cite{dbrane}.
In this model, scalar fermion masses are not completely universal and
gaugino mass unification can be relaxed. Also the string scale is around
$10^{12}$ GeV (the intermediate scale) rather than GUT scale.

Since there are now three moduli ($T_i$)  and one dilaton superfields 
in this case, we use the following parametrization that is appropriate for 
several $T_i$ moduli:
\begin{eqnarray}
  F^S & = & \sqrt{3}~( S + S^* )~ m_{3/2} \sin\theta,
\nonumber   \\
  F^i & = & \sqrt{3}~( T_i + T_i^* )~ m_{3/2} \cos\theta~\Theta_i
\end{eqnarray}
where $\theta$ and $\Theta_i ~(i = 1,2,3)$ with $\sum_i | \Theta_i |^2 = 1$
parametrize the directions of the goldstinos in the $S, T_i$ field space.
%
The explicit expressions for the soft terms 
are given in Ref.~\cite{dbrane}. 
Let us simply note that 
the scalar and the gaugino masses become nonuniversal for generic 
goldstino angles, and there could be larger flavor violations 
in the low energy processes as well as
enhanced SUSY contributions to the $a_\mu^{\rm SUSY}$.

Therefore the $D$ brane model considered in this work is specified by
following six parameters :
\[
m_{3/2},~~\tan\beta,~~\theta,~~\Theta_{i=1,2},~~ {\rm sign} (\mu).
\]
Earlier phenomenological analysis of $D$ brane models can be found on 
the muon $(g-2)_\mu$~\cite{munoz2}.  
The discussion on  $B\rightarrow X_s \gamma$, $B\rightarrow X_s l^+ l^-$ 
and $B_s \rightarrow \mu^+ \mu^-$ in this scenario is given in 
Ref.s~\cite{Baek:2002wm}. 
Here, we combine the DM scattering and the branching ratio for 
$B_s \rightarrow \mu^+ \mu^-$. 
We fix $\tan\beta = 50$ and scan over the following parameter space :
$-\pi/4 \leq \theta \leq \pi /4$, $m_{3/2} \leq 1000$ GeV, and
$\Theta_i $ 
in order to search the allowed parameter space. 
In this scenario again, it turns out that the current upper limit on 
$B( B_s \rightarrow \mu^+ \mu^- )$ already puts a strong constraint 
on the parameter space in the $D-$brane scenarios. 
In Fig.~5 (a), we show the correlation between 
$B( B_s \rightarrow \mu^+ \mu^- )$ and $\sigma_{\chi p}$. 
In Fig.~5 (b), we show the DM cross section as a function of the LSP
mass $m_{\chi}$. Note that the upper limit on  
$B( B_s \rightarrow \mu^+ \mu^- )$ makes a stringent constraint on the
model, especially for light LSP mass $m_{\chi} \lesssim 150$ GeV.   
If we ignored the upper
limit on $B( B_s \rightarrow \mu^+ \mu^- )$, then the resulting DM 
scattering cross section could be well within the CDMS region with 
$ \sigma_{\tilde{\chi} p} > 4 \times 10^{-7}$ pb.  
However, such a large DM scattering cross section implies too large a 
branching ratio for 
$B ( B_s \rightarrow \mu^+ \mu^- ) > 5.8 \times 10^{-7}$ for light LSP
$m_{\chi} \lesssim 150$ GeV, and thus has to be discarded. 
For heavier LSP mass, both constraints have to be considered altogether,
since they are complementary to each other. 

\begin{figure*}
\begin{tabular}{cc}
{\includegraphics[height=5.0cm]{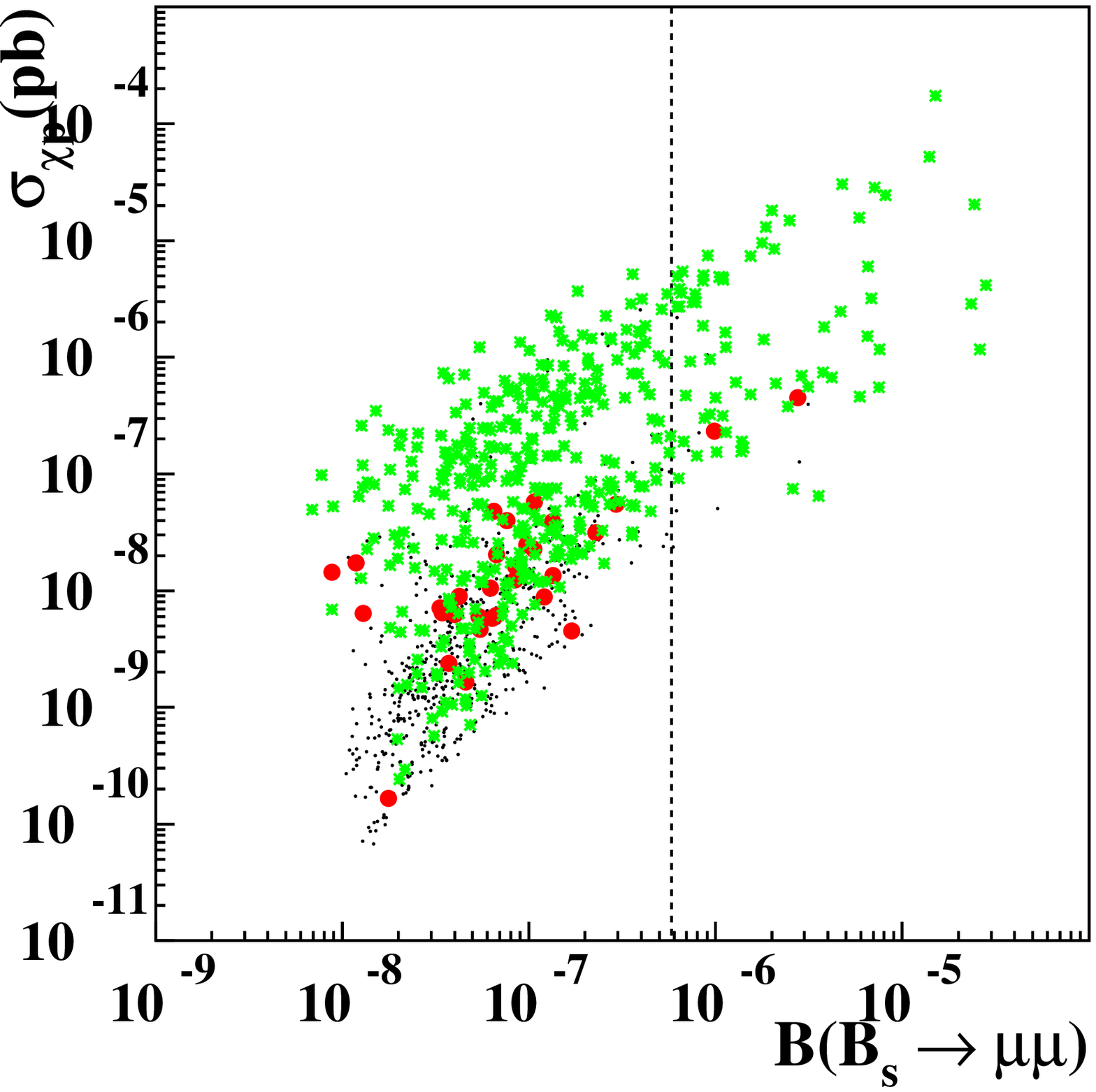}}&
{\includegraphics[height=5.0cm]{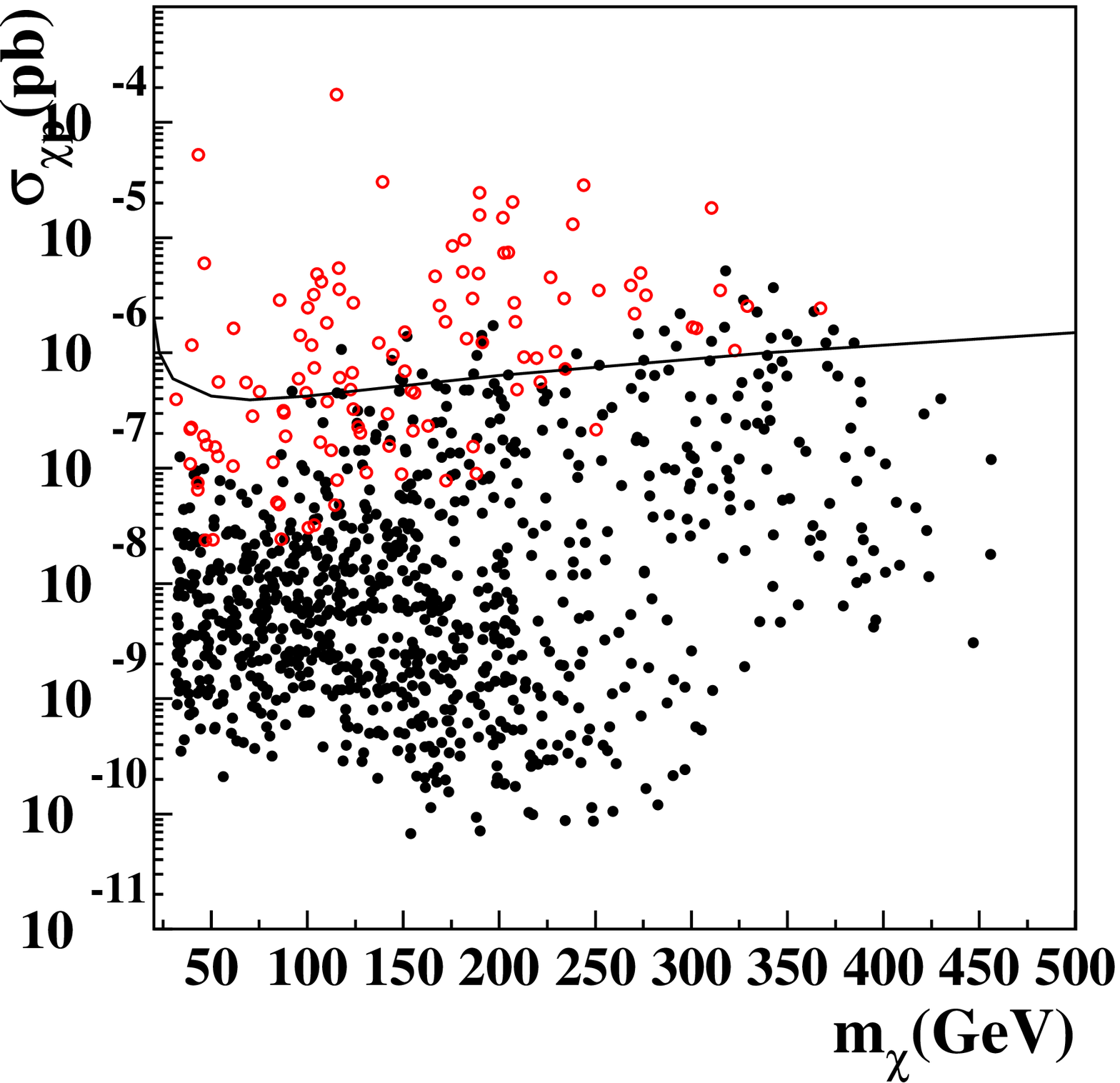}}
\end{tabular}
\caption{\label{fig4} 
$\sigma_{\tilde{\chi} p}$ vs. $B( B_s \rightarrow \mu^+ \mu^- )$ within 
$D-$brane models of Ref.~\cite{dbrane}. 
}
\end{figure*}

\begin{figure}
\includegraphics[height=5.0cm]{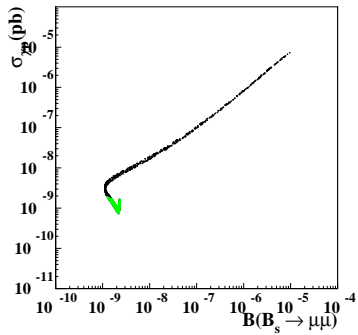}%
\caption{\label{fig5} 
$\sigma_{\tilde{\chi} p}$ vs. $B( B_s \rightarrow \mu^+ \mu^- )$ within 
the AMSB model with $M_{\rm aux} = 50$ TeV. Black dots are excluded by the
upper limit on $B \rightarrow X_s \gamma$ branching ratio, whereas the green
dots satisfy all the constraints.
}
\end{figure}

The DM scattering in the AMSB scenarios is qualitatively similar to the 
previous cases. Although the LSP in the AMSB scenarios is winolike in this 
case, Higgs boson contribution to the DM scattering is still important.
And there is a strong correlation between $\sigma_{\tilde{\chi} p}$ 
and $B ( B_s \rightarrow \mu^+ \mu^- )$.  
In the simplest version of the AMSB model, one adds a common scalar mass 
$m_0^2$ to scalar mass parameters in order to evade the tachyonic slepton 
mass problem. In Fig.~\ref{fig5}, we show the scattered plot for 
$\sigma_{\tilde{\chi} p}$ and $B ( B_s \rightarrow \mu^+ \mu^- )$ within 
such an AMSB scenario with $M_{\rm aux} = 50$ TeV. The black dots are 
excluded by $B\rightarrow X_s \gamma$ constraint, and only the green points
survive.  The resulting predictions for the DM scattering and the branching 
ratio for $B_s \rightarrow \mu^+ \mu^-$ is so small that this class of 
the AMSB model has no observable effects in the DM scattering or 
$B_s \rightarrow \mu^+ \mu^-$.

In the heterotic $M$ theory of Horava and Witten,  we have the similar 
correlation between $\sigma_{\tilde{\chi} p}$ and 
$B ( B_s \rightarrow \mu^+ \mu^- )$ in the large $\tan\beta$ region.  
However, after imposing direct search bounds on Higgs and SUSY particle 
masses as well as $B\rightarrow X_s \gamma$ constraint and the neutralino
LSP condition,  the resulting DM scattering cross section turns out 
very small: $\sigma_{\tilde{\chi} p} \lesssim 10^{-8}$ pb, which is 
well below the sensitivity of the current DM search experiments. .
Also we get $B ( B_s \rightarrow \mu^+ \mu^- ) < 10^{-7}$ which is 
beyond the reach of Tevatron Run II. 


In conclusion, we pointed out that there is a strong correlation between 
the neutralino dark matter scattering cross section with nuclei 
and the branching ratio for $B_s \rightarrow \mu^+ \mu^-$ 
within a large  class of supergravity models. This correlation arises 
mainly from $\tan\beta$ and heavy neutral Higgs masses $(m_H , m_A )$.
We have discussed mSUGRA with (non)universal gaugino masses and 
(non)universal scalar masses and supergravity scenarios derived from 
heterotic $M$ theory, and AMSB scenario. In the $D$ brane scenario 
considered in this work, the correlation is diluted because of nonuniversal
scalar and gaugino mass parameters. Still the upper limit on 
$B ( B_s \rightarrow \mu^+ \mu^- )$ puts a very strong constraint 
on DM cross section, even stronger than the CDMS limit. 
Thus the decay $B_s \rightarrow \mu^+ \mu^-$ could give invaluable 
informations not only on  SUSY breaking mediation mechanisms as noticed in
Refs.~\cite{Baek:2002rt,Baek:2002wm}, but also give a strong constraint 
on the neutralino 
DM scattering cross section within a large class of supergravity models
in the large $\tan\beta$ region.  
This is another interesting example of complementarity between rare 
$B_s$ decays (indirect probe of SUSY) and DM scattering 
(direct probe of SUSY). 

\begin{acknowledgments}
This work is supported in part by KOSEF Sundo grant R02-2003-000-10085-0,
KOSEF through CHEP at Kyungpook National
University, KRF grant KRF-2002-070-C00022, and by BK21 Haeksim program.
The work of YGK was supported by the Korean Federation of Science and 
Technology Societies through the Brain Pool program.
\end{acknowledgments}


\end{document}